\documentclass[prl,preprint,amsmath,amssymb,showpacs,showkeys,superscriptaddress]{revtex4}
\usepackage{graphicx}
\usepackage{dcolumn}
\newcommand{\be}{\begin{equation}}
\newcommand{\ee}{\end{equation}}
\newcommand{\ba}{\begin{eqnarray}}
\newcommand{\ea}{\end{eqnarray}}

\begin{document}

\title{Thermodynamic stability of fluid-fluid phase separation in binary athermal mixtures: The role of nonadditivity}

\author{G. Pellicane}
\altaffiliation{Corresponding Author} \email{\tt
giuseppe.pellicane@unime.it} \affiliation{Universit\`a degli Studi
di Messina, Dipartimento di Fisica \\ Contrada Papardo, 98166
Messina, Italy}
\author{F. Saija}
\email{\tt saija@me.cnr.it}
\affiliation{CNR - Istituto per i Processi Chimico-Fisici,
Sezione di Messina,\\Via La Farina 237, 98123 Messina, Italy}
\author{C. Caccamo}
\email{\tt carlo.caccamo@unime.it} \affiliation{Universit\`a degli
Studi di Messina, Dipartimento di Fisica \\ Contrada Papardo,
98166 Messina, Italy}
\author{P. V. Giaquinta}
\email{\tt paolo.giaquinta@unime.it} \affiliation{Universit\`a
degli Studi di Messina, Dipartimento di Fisica \\ Contrada
Papardo, 98166 Messina, Italy}

\date{\today}

\begin{abstract}
We study the thermodynamic stability of fluid-fluid phase
separation in binary nonadditive mixtures of hard-spheres for
moderate size ratios. We are interested in elucidating the role
played by small amounts of nonadditivity in determining the
stability of fluid-fluid phase separation with respect to the
fluid-solid phase transition. The demixing curves are built in the
framework of the modified-hypernetted chain and of the
Rogers-Young integral equation theories through the calculation of
the Gibbs free energy. We also evaluate fluid-fluid phase
equilibria within a first-order thermodynamic perturbation theory
applied to an effective one-component potential obtained by
integrating out the degrees of freedom of the small spheres. A
qualitative agreement emerges between the two different
approaches. We also address the determination of the freezing line
 by applying the first-order thermodynamic perturbation
theory to the effective interaction between large spheres. Our
results suggest that for intermediate size ratios a modest amount
of nonadditivity, smaller than earlier thought, can be sufficient
to drive the fluid-fluid critical point into the thermodinamically
stable region of the phase diagram. These findings could be
significant for rare-gas mixtures in extreme pressure and
temperature conditions, where nonadditivity is expected to be
rather small.
\end{abstract}
%\pacs{}
%\keywords{}
\maketitle

\section{Introduction}
Mixtures of large and small colloidal particles can be modelled
with a binary hard-sphere mixture. Depending on the value of the
size ratio $y=\sigma_2/\sigma_1$, where $\sigma_1$ and $\sigma_2$
are the large and small sphere diameters, respectively, and on the
packing fractions of the two species $\eta_1$ and $\eta_2$, the
small spheres induce effective attractions among large particles.
For size ratios very different from one, the strength of these
attractions can be such as to originate a fluid-fluid phase
separation (FFPS). In fact, several studies show that phase
separation switches on for very asymmetric size ratios, i.e.
$y\approx 0.1$ (see Ref. \cite{dijk1} and references therein) and
that it is always preempted by the freezing transition. The
physical origin of this attraction is found in the osmotic
depletion effect, i.e., in the gain of free volume available to
the small spheres due to the overlap of excluded volumes of
approaching larger spheres. On the other hand, a more efficient
packing for phases involving a majority of particles of the same
species can be obtained also by means of a positive nonadditivity
$\Delta$ in the cross interaction diameter
$\sigma_{12}=1/2(\sigma_1+\sigma_2)(1+\Delta)$ \cite{dijk2}. In
real sytems, such as sterically or electrostatically stabilized
colloids, nonadditivity is expected to be rather small
\cite{louis}. Nonadditivity should be a persistent feature of
colloidal systems associated with the presence of small residual
interactions that cannot be modelled through a hard-sphere
potential and it is expected to have a very pronounced effect on
the phase behavior \cite{louis}. For instance, assuming that
$\sigma_2$ is the radius of gyration of a nonadsorbing polymer,
experiments on colloid-polymer mixtures indicate that a
fluid-fluid demixing transition develops for $y \ge 0.35$
\cite{lekker}, while no fluid-fluid demixing has been reported in
literature for additive hard-sphere (AHS) mixtures such size
ratios \cite{dijk1}. Another class of colloidal systems for which
nonadditivity effects can be profitably introduced includes
surfactants; in fact, nonadditive interactions between hard-chain
models of surfactants and the solvent allow to study more
efficiently common self-assembly mechanisms as, for instance,
micelle or reverse micelle formation and double layer structures
\cite{abu}. Finally, deviations from the usual Lorentz-Berthelot
mixing rules are observed also in rare-gas mixtures \cite{rowli},
expecially in extreme thermodynamic regimes \cite{barrat}. For
instance, nonadditivities lower than $0.1$ can be used to model
$H_2$-rare gas mixtures, in order to account for the
extra-attractive interactions between hydrogen molecules, due to
electron exchange at very high pressures \cite{vanderberg}.

In general, the depletion effect in AHS mixtures is stronger for
smaller size ratios while nonadditivity is expected to enhance the
homocoordination more efficiently when the two species become more
similar. The competition between these two different mechanisms,
in the fulfilement of the conditions of maximum entropy, suggests
that for intermediate size ratios even a small amount of
nonadditivity can lead to a thermodynamically stable phase
separation. This picture is confirmed on the microscopic side: in
fact, the range of the effective depletion potential for the large
spheres becomes very small for large size-asymmetries (being of
the order of magnitude of the diameter of the small spheres) and
the correlation-induced repulsive barrier becomes wider for small
size-asymmetries \cite{louis,louis2}. Obviously, these two effects
favour the metastability of liquid-vapor equilibrium, that can be
mapped onto the fluid-fluid equilibrium of the mixture as, we
shall see later on; if we consider that, on increasing $\Delta$,
the attractive well of the effective potential becomes deeper
while the repulsive barrier remains roughly the same, we conclude
again that the optimal conditions to stabilize a FFPS with a small
amount of nonadditivity could be met for intermediate
size-asymmetries.

Recently, Lo Verso and coworkers employed the effective potentials
previously obtained by other authors to investigate the phase
diagram of very asymmetric nonadditive mixture by means of the
hierarchical reference theory \cite{loverso}. However, there is
little and definitely not exaustive information about the
fluid-fluid phase stability of nonadditive hard-sphere (NAHS)
mixtures in the range of relatively small (y $\approx$ $0.8$) to
intermediate (y $\approx$ $0.3$) size-asymmetries. Moreover, there
is a strong evidence that a stable FFPS cannot be achieved with a
small amount of nonadditivity for large size-asymmetries. For
instance, Louis and coworkers \cite{louis2} found that $\Delta$
should be $\approx$ $0.2$ for y=$0.2$ in order to stabilize the
FFPS. This happenstance should persist also for larger
size-asymmetries because the gap of packing fractions between the
critical point and the freezing line remains almost unchanged up
to y=$0.033$ for AHS mixtures \cite{dijk1}.

Altough the qualitative phase behavior of hard-sphere mixtures
 is known, a few investigations have been carried out by
means of integral-equation approaches. In this paper we study the
phase diagram of NAHS mixtures for size ratios
y=$0.75$,$0.6$,$0.5$,$0.3$, and for nonadditivities $\Delta =
0.05,0.1$. We build the FFPS curves according to the following
procedures: We perform extensive modified-hypernetted chain (MHNC)
\cite{mhnc} and Roger-Young (RY) \cite{ry} Gibbs free energy
calculations on the full mixture. We also build the FFPS curve by
exploiting a first-order perturbation theory \cite{hansen} in a
crude representation of the effective depletion potential for the
larger spheres in contact with a reservoir of small spheres at a
packing fraction $\eta_s^r$. The resulting potential was obtained
in the framework of the Derjaguin approximation by Gotzelmann and
coworkers \cite{gotz} for AHS mixtures and extended to the
nonadditive case by Louis and coworkers \cite{louis}. Then, we map
the liquid-vapor phase diagram of the effective depletion
potential onto the fluid-fluid phase diagram of the binary mixture
according to an explicit conversion formula, based on Rosenfeld's
fundamental measure density-functional theory (DFT) \cite{rosen},
between $\eta_s$ - the packing fraction of the small spheres of
the system - and the packing fraction in the reservoir $\eta_s^r$
\cite{roth}. Similarly, the calculation of the freezing line is
performed by mapping the freezing line of the effective potential
onto the freezing line of the binary mixture.

The paper is organized as follows: In Section II, we present the
models and the theoretical procedures, while in Section III we
show the results for the phase diagrams for different size ratios.
Finally, we give some concluding remarks in Section IV.

\section{Model and methods}
We consider a binary system of particles
interacting through the pair potential

\begin{equation}
v_{ij}(r) = \left\{ \begin{array}{ll}
+\infty & \textrm{$r<\sigma_{ij}$} \\
0 & \textrm{$r\geq\sigma_{ij}$}
\end{array} \right. i,j = 1, 2
\label{potij}
\end{equation}

\noindent where $\sigma_{i}$ is the diameter of the i-th species,
and $\sigma_{ij}= 1/2 {( \sigma_{ii} + \sigma_{jj} ) (1 +
\Delta)}$ . The mixture can be described by the size ratio $y$, by
the partial and total number density of particles $\rho_i$ and
$\rho$, respectively, and by the mole fraction of species $i$,
 $x_{i}=\frac{\rho_i}{\rho}$.

The Ornstein-Zernike (OZ) equations \cite{report} for the
homogeneous mixture are:
\begin{equation}
h_{ij}(r) = c_{ij}(r) + \sum^{2}_{k=1} \rho_k \int c_{ik} (|{\bf
r} - {\bf r'}|) h_{kj} ({\bf r'}) d {\bf r'}, \label{hij}
\end{equation}

\noindent where $h_{ij}(r) =g_{ij}(r)-1$ and $c_{ij}(r)$ are the
pair correlation and the direct correlation function,
respectively.

We solved the OZ equations under the MHNC closure \cite{mhnc},
based on the exact relationship obtained through cluster expansion
techniques:

\begin{equation}
g_{ij}(r) = \exp [-\beta v_{ij}(r) + h_{ij}(r) - c_{ij}(r)
                                   + E_{ij}(r)] \quad .
\end{equation}

\noindent We approximated the bridge functions $E_{ij}(r)$ with
their Percus-Yevick \cite{py} counterparts for hard sphere
mixtures, $E^{HS}_{ij}(r;\sigma^*_{ij})$. The parameters
$\sigma^*_{ij}$ are used to ensure thermodynamic self-consistency.

We also solved the OZ equation in the RY \cite{ry} approximation:

\begin{equation}
g_{ij}(r) = \exp [-\beta v_{ij}(r)] \: \left\{1
 + \frac{ \exp \:\{\:f_{ij}(r) \: [\: h_{ij}(r)
-c_{ij}(r)\: ]\:\} - 1 } {f_{ij}(r)}\right\}, \label{gijry}
\end{equation}

\bigskip

\noindent where $f_{ij}(r) = 1 - \exp[\:\xi_{ij}r\:]$ and the
quantities $\xi_{ij}$ are adjusted in such a way so to satisfy the
thermodynamic consistency of the theory.

In order to ensure the internal thermodynamic consistency of the
theories we equated the two osmotic compressibilities evaluated by
differentiating the virial pressure~\cite{hansen,report}

\begin{equation}
\left (\frac{\beta P}{\rho} \right )^{vir} ={\frac{ 2\pi}{3} }
\rho \sum_{ij} {x_{i}x_{j}\sigma^3_{ij}g_{ij}(\sigma_{ij}) },
\label{virial}
\end{equation}

\noindent where $\beta = 1/k_{B}T$ and $k_{B}$ is the Boltzmann
costant, and the compressibilities resulting from the fluctuation
theory ~\cite{report}:

\begin{equation}
1 - \sum_j \rho_j \tilde{c}_{ij} (q=0) = \left ( \beta {\frac{\partial P}
{\partial \rho_i}} \right )^{vir}_{T, \rho_{j (j \neq i)}},
\label{comp}
\end{equation}
\noindent where $\tilde{c}_{ij}(q)$ is the Fourier transform of
$c_{ij}(r)$ and $i =1,2$.

\begin{figure}
\begin{center}
\includegraphics[width=9cm,angle=0]{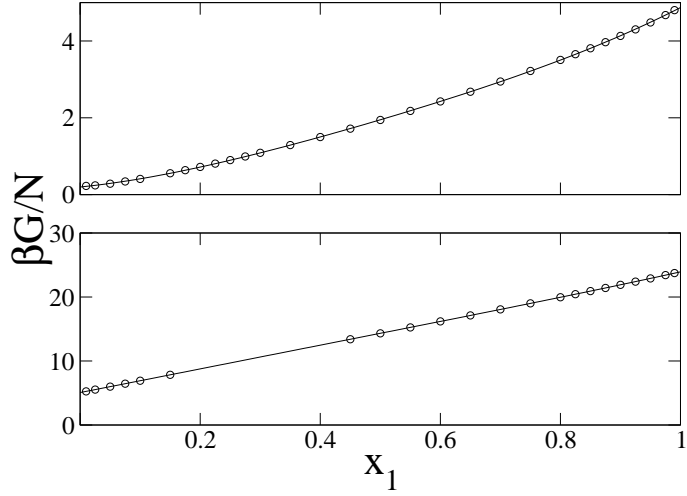}
\end{center}
\caption{Total Gibbs free energy versus mole fraction for $y=0.5$,
$\Delta=0.05$ at a pressure lower ($P^*=0.5$, upper panel) and higher
($P^*=2.74$, lower panel) than the critical one. Circles are  calculated
as a sum of Eq. (\ref{gex}) and of Eq. (\ref{gid}), while the full line
is a polynomial of the fourth order interpolating the data.} \label{fig. 1}
\end{figure}

\begin{figure}
\begin{center}
\includegraphics[width=9cm,angle=0]{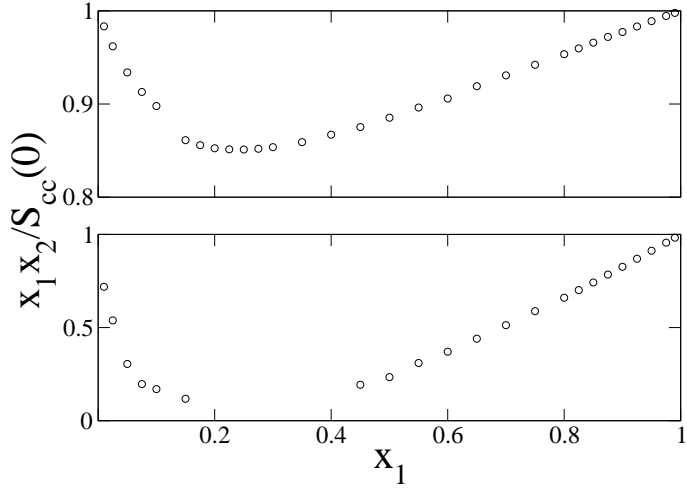}
\end{center}
\caption{Inverse ratio between the concentration-concentration
structure factor at zero wave vector $S_{cc}(q=0)$ and the
corresponding value for an ideal mixture versus mole fraction for
$y=0.5$, $\Delta=0.05$ at a pressure lower
 ($P^*=0.5$, upper panel) and higher ($P^*=2.74$, lower panel) than the critical one.} \label{fig. 2}
\end{figure}

We considered at least thirty mole fraction values and evaluated
the total Gibbs free energy from the excess Helmholtz free energy
per particle, that was obtained by integrating the excess part of
the virial pressure (Eq.\ (\ref{virial})) as a function of the
total density of the mixture. We obtained the  excess contribution
to the Gibbs free energy \cite{hansen,bijor} as:

\begin{equation}
\frac{\beta G^{ex}}{N} = \frac{\beta A^{ex}}{N} + Z - 1 - \ln{Z};
\label{gex}
\end{equation}

\noindent then, we calculated the total Gibbs free energy by
adding the ideal part (we omit the kinetic part associated with
the de Broglie wavelength)

\begin{equation}
\frac{\beta G^{id}}{N} = \ln{Z} + \sum_{i}^{} x_{i}\ln{\rho_{i}} =
\ln{\beta P} + \sum_{i}^{} x_{i}\ln{x_{i}} \label{gid}
\end{equation}

\noindent where $Z = \beta P / \rho$. The Gibbs free energy was
interpolated for each mole fraction with cubic splines, and these
fits were used to determine the Gibbs free energy at costant
pressure. Finally, the FFPS was obtained by applying the
construction of the common tangent to this latter quantity. By
numerical inspection we found that the Gibbs free energy plotted
as a function of the concentration at constant pressure can be
interpolated by a quartic polynomial. On the other hand, for
pressures greater than the critical one, the Gibbs free energy
turns out to be sampled just on a limited number of state points.
This drawback is due to the loss of stability in proximity of the
fluid-fluid spinodal of the numerical alghoritm used to solve
integral equation theories \cite{gillan}. Polynomial fits of a
small number of data may bias the estimate of the FFPS; however,
we noted that the mixture separates into two different equilibrium
compositions when the discriminant of the second-order equation
(the second derivative of the Gibbs free energy) attains a value
greater than zero. Thus, in order to guarantee the safety of the
overall procedure, we monitored the trend of the discriminant as a
function of the total pressure from negative values (at which the
Gibbs free energy is fully sampled on a grid of 30 points because
there is no phase coexistence) up to positive values. We trusted
the calculation of the coexistence concentrations in the pressure
range where the discriminant varies without manifest
discontinuities; in fact, jumps of the discriminant plotted as a
function of the pressure are trivially related to a bias
introduced by the poor sampling of the Gibbs free energy. We show
in Figure \ref{fig. 1} the typical shape of the Gibbs free energy
for $y=0.6$ and $\Delta=0.05$, corresponding to a subcritical
pressure $P^*=P\sigma^3/\epsilon=0.5$ (see upper panel) and to a
supercritical pressure $P^*=2.74$ (see lower panel). In Figure
\ref{fig. 2} we report, for the same pressures, the inverse ratio
between the concentration-concentration structure factor at zero
wave vector $S_{cc}(q=0)$ and the corresponding value for an ideal
mixture, a quantity that provides a measure of how much the
numerical procedure allows one to to approach the fluid-fluid
spinodal.

\begin{figure}
\begin{center}
\includegraphics[width=9cm,angle=0]{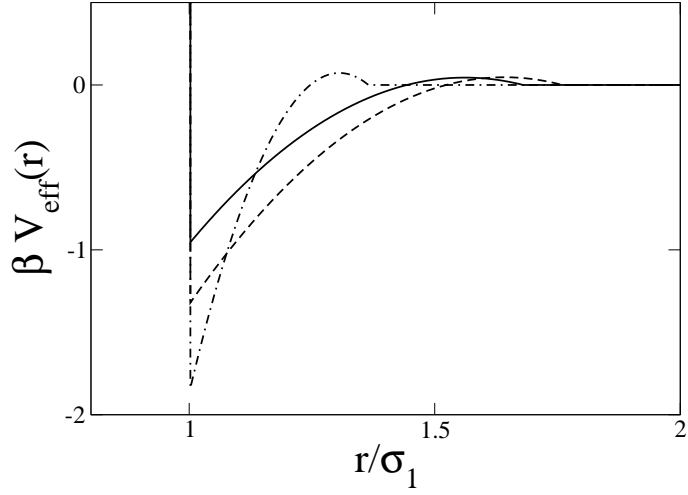}
\end{center}
\caption{Effective potentials of Eq. (\ref{eq4}) for
$\eta_2^r=0.15$: $y=0.6$, $\Delta=0.05$ (full line), $y=0.6$,
$\Delta=0.1$ (dashed line), $y=0.3$, $\Delta=0.05$ (dot-dashed
line).} \label{fig. 3}
\end{figure}

The FFPS was also calculated according to the following procedure:
We considered a simple form of the depletion potential for the
larger spheres in contact with a reservoir of small spheres with
packing fraction $\eta_2^r$ \cite{louis2,gotz}:

\begin{equation}
\beta V_{eff}(r) = \left\{
\begin{array}{lll}
\infty & $r$\leq\sigma_{1}
\nonumber \\
\frac{-3 \eta_{2}^{r}(1+y_{eff})}{2 y^{3}}
\biggl\{
h(r)^2+\eta_2^r
\left[4 h(r)^{2} -3 y h(r) \right] \nonumber \\
+ (\eta_{2}^{r})^{2} \left[10 h(r)^{2} - 12 y h(r) \right]
\biggl\}
& \sigma_{1}\leq $r$\leq\sigma_{1}(1 + y_{eff})
\nonumber \\
0 & $r$\geq\sigma_{1}(1 + y_{eff})
\end{array} \right.
\label{eq4}
\end{equation}
where the effective size ratio is:
$$
y_{eff} = y + \Delta + \Delta y,
$$
while $h(r) = (1 + y_{eff}) - r/\sigma_1$. The one-component
potential of interaction is shown in Figure \ref{fig. 3} for some
different parameters. Dijkstra and coworkers \cite{dijk1} have
shown that this simple form for the effective interaction between
larger spheres is reliable also on approaching $y$=$1$. We built
the liquid-vapor phase diagram of the depletion potential by means
of a first-order thermodynamic perturbation theory (see Refs.
\cite{louis, costa} for details) by equating the chemical
potentials of the two phases at the same pressure. Then, the
reservoir packing fraction $\eta_s^r$ was converted into the
packing fraction of the smaller spheres in the real mixture by
means of the relation:
\begin{eqnarray} \label{etas}
\eta_s & = & (1-\eta_b) \eta_s^r - 3 y_{eff}~ \eta_b \eta_s^r
\frac{(1-\eta_s^r)}{(1+2\eta_s^r)} \nonumber\\
& & - 3 y_{eff}^2~ \eta_b \eta_s^r \frac{(1-\eta_s^r)^2}{(1+2
\eta_s^r)^2}
 - y_{eff}^3~ \eta_b \eta_s^r
\frac{(1-\eta_s^r)^3}{(1+2 \eta_s^r)^2},
\end{eqnarray}
\noindent so that the liquid-vapor phase coexistence can be mapped
onto the FFPS for the mixture. This relation has been successfully
tested against simulation results in the grand-canonical ensemble
by other authors \cite{roth}.

As far as the freezing line is concerned, while in the considered
range of size-asimmetries ($y$ between $0.3$ and $0.75$) no
substitutionally disordered crystals are expected to be
thermodynamically stable, superlattice structures can be found for
additive mixtures with stoichometric compositions ($AB_2$,
$AB_{13}$) when $y$ $\leq$ $0.6$ \cite{cottin,hunt}. However, the
determination of the phase stability of a binary superlattice
depends on three variables, i.e. the size ratio $y$, the total
packing fraction $\eta$, and the mole fraction of one of two
species, say $x_1$. This calculation represents a formidable task.
The situation is even worse for NAHS mixtures because of the
additional variable $\Delta$. On the other hand, it has been
conjectured \cite{louis2} that, at least for sufficiently small
values of $y$ and for moderate densities, the larger spheres might
form a face-centered cubic (FCC) lattice permeated by a fluid of
small spheres; thus, in the present paper we limited our
investigation to the determination of the freezing line to this
crystalline simmetry. Again, we resorted to a first-order
perturbation theory applied to the effective depletion potential
in order to determine the chemical potentials both in the fluid
and in the solid phase \cite{louis}. Once the freezing line for
the effective potential is determined, by equating the chemical
potentials at costant pressure, the freezing line of the full
mixture can be reconstructed by means of Eq. (\ref{etas}).

\section{Results and discussion}

\begin{figure}
\begin{center}
\includegraphics[width=9cm,angle=-90]{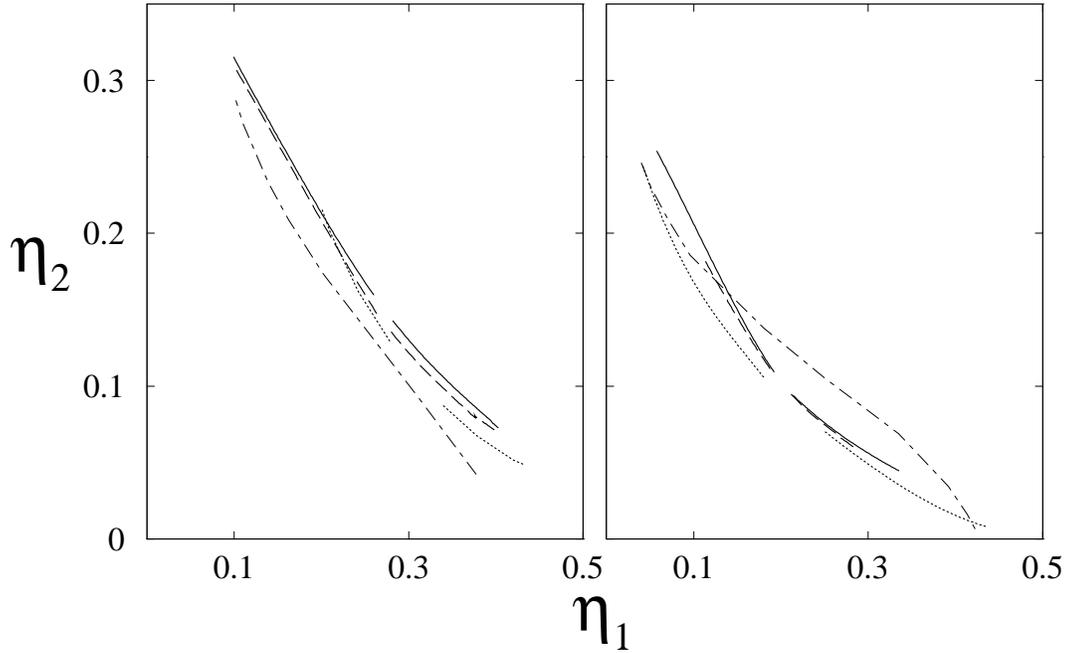}
\end{center}
\caption{Fluid-fluid and fluid-solid transition lines in the
$\eta_2$ vs. $\eta_1$ representation for $y=0.75$, $\Delta=0.05$
(left panel) and $\Delta=0.1$ (right panel). Fluid-fluid phase
coexistence. Solid line: MHNC; dashed line: RY; dotted line:
first-order perturbation theory (PT) on the effective potential.
Freezing line. Dot-dashed line: first-order perturbation theory on
the effective potential.} \label{fig. 4}
\end{figure}

We start our analysis of the results obtained for the largest size
ratio considered, i.e., $y=0.75$. The FFPS curves calculated from
integral equation theories for $\Delta = 0.05$ (see full and
dotted line of the left panel of Figure \ref{fig. 4}) turned out
to be in good agreement between each other. The critical points
were located in the region between the two branches of coexisting
phases, where no curve was reported in the figure because of
numerical problems arising from the proximity of the critical
point. In general, we found a good agreement between RY and MHNC
estimates of the critical parameters of the FFPS also for all the
other case-studies considered in this work (see Table \ref{t2}).
Moreover, we found that the locus of points where the residual
multi-particle entropy vanishes for $y=0.75$ \cite{saija},
resembling the spinodal curve, is congruent with the FFPS curves
obtained in this work with Gibbs free energy calculations. The
freezing line (see the dash-dotted line) was located below the
critical points, so that these FFPSs are thermodynamically
metastable. The FFPS calculated from the depletion potential (see
dotted line) was in qualitative agreement with the FFPS calculated
from integral equation theories; in fact, the FFPS estimated from
the depletion potential was metastable with respect to the
freezing transition as well, because its critical parameters fall
above the freezing line. It is worth noting that Louis and
coworkers \cite{louis} found a good agreement between the
first-order perturbation theory and the computer simulation as
applied to the description of the fluid-solid equilibrium, with a
slight tendency of the theoretical freezing line to underestimate
the simulation one; this drawback of the perturbation theory
suggests that, in principle, it could be possible to have a stable
fluid-fluid critical point even for a $\Delta$ value as low as
$0.05$; this is particularly true for the size ratio $y=0.75$
because the gap of packing fractions between the FFPS critical
point and the freezing line turned out to be lower, as we shall
see later on.

A better agreement between the critical parameters calculated
through integral equation theories and the perturbation theory was
found for the largest nonadditivity $\Delta = 0.1$ (see the right
panel reported in Figure \ref{fig. 4}). Upon doubling the
nonadditivity parameter from $0.05$ to $0.1$, the critical
concentration $x_1$ only slightly changed. More interestingly, all
the theoretical critical parameters fall below the freezing line
(see dash-dotted line), so that all the predicted FFPS were
thermodynamically stable with respect to the freezing transition
for $\Delta = 0.1$.

\begin{figure}
\begin{center}
\includegraphics[width=9cm,angle=-90]{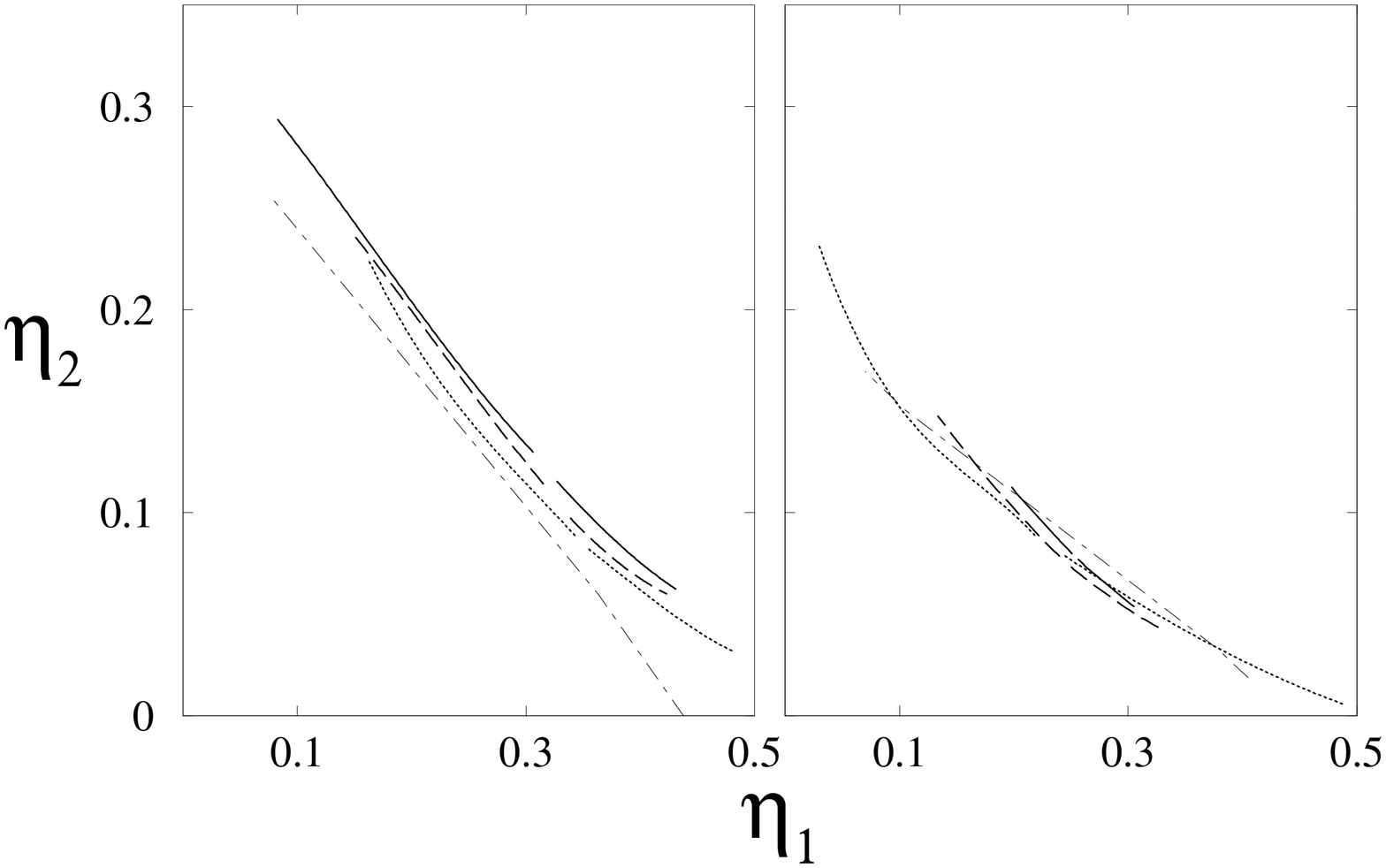}
\end{center}
\caption{Fluid-fluid and fluid-solid transition lines in the
$\eta_2$ vs. $\eta_1$ representation for $y=0.6$, $\Delta=0.05$
(left panel) and $\Delta=0.1$ (right panel). See Figure
 \ref{fig. 4} for the legend.} \label{fig. 5}
\end{figure}

\begin{figure}
\begin{center}
\includegraphics[width=9cm,angle=-90]{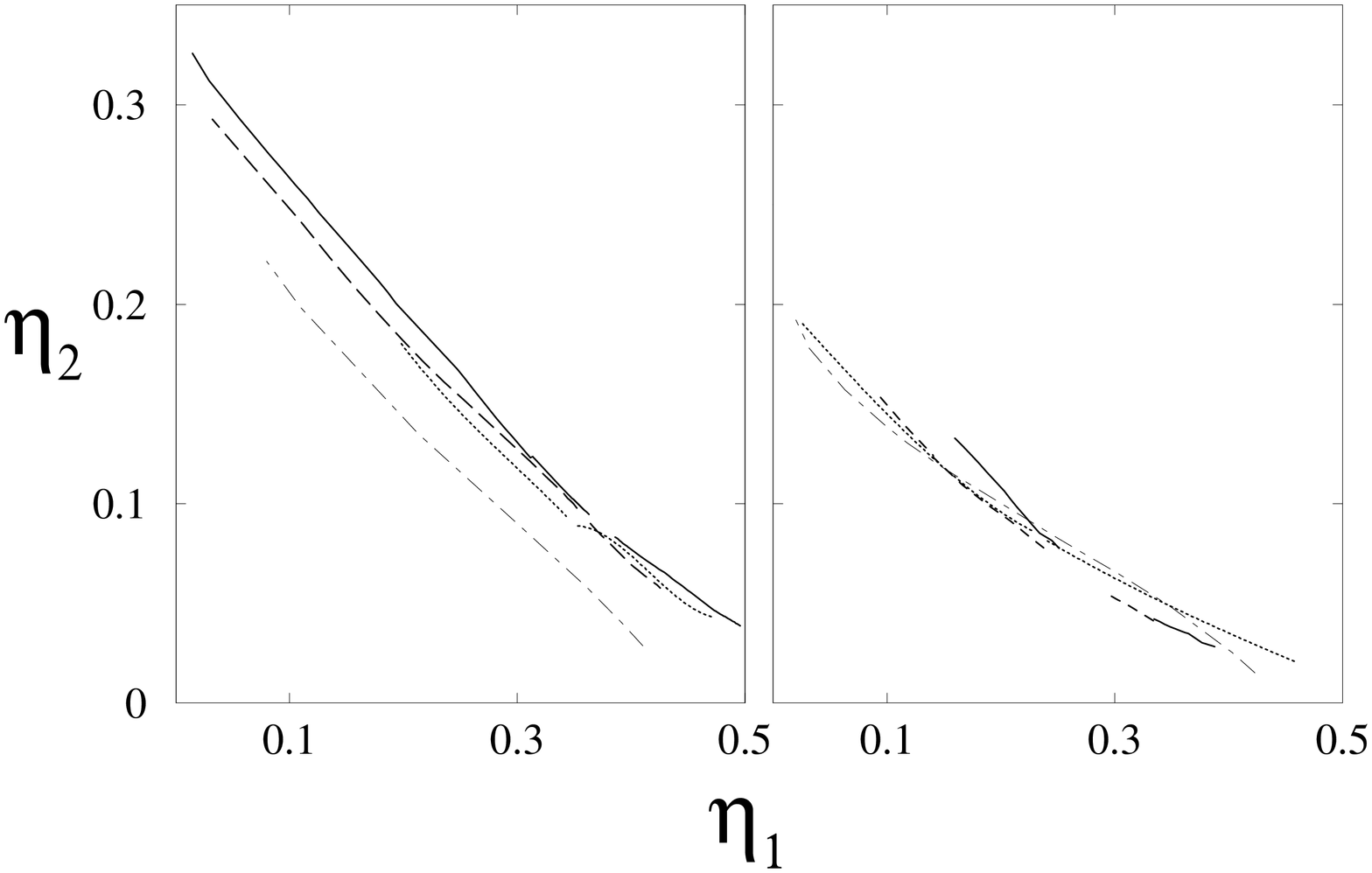}
\end{center}
\caption{Fluid-fluid and fluid-solid transition lines in the
$\eta_2$ vs. $\eta_1$ representation for $y=0.5$, $\Delta=0.05$
(left panel) and $\Delta=0.1$ (right panel). See Figure
 \ref{fig. 4} for the legend.} \label{fig. 6}
\end{figure}

The phase diagrams obtained for size ratios $y=0.6$ and $y=0.5$,
which are reported in Figures \ref{fig. 5} and \ref{fig. 6},
respectively, are particularly meaningful for Helium-Xenon
mixtures at high pressure. In fact, some authors \cite{barrat}
estimate that the size ratio for such systems could lie between
$0.6$ and $0.5$, depending on the choice of the distance of
impenetrability of the atoms. For $\Delta=0.05$ some discrepancies
between the FFPS curves calculated within the two integral
equation theories (IETs) emerged, but the critical parameters
turned out to be very similar (see Table \ref{t2}). The FFPS
critical parameters calculated from the depletion potential (see
dotted line of Figure \ref{fig. 5} and \ref{fig. 6}) were in
fairly good agreement with the integral equation results.

For $\Delta = 0.1$ all the theoretical results suggested that the
FFPS become thermodynamically stable, as already observed for the
case $y=0.75$. However, the trend observed starting from $y=0.75$
up to $y=0.5$ for $\Delta = 0.1$, suggested that a size ratio
close to $y=0.5$ could constitute a lower bound for the
thermodynamic stability of the fluid-fluid phase separation; in
fact, the gap of packing fractions between the FFPS critical point
and the freezing line at the critical concentration shrank in
passing from $y=0.75$ to $y=0.5$ and became very narrow for this
latter size ratio. We experienced some difficulties in estimating
the FFPS in the framework of the IETs for $y=0.5$, $\Delta = 0.1$,
so that the critical parameters are affected by a large error, as
illustrated in Table \ref{t2}.

The smallest size ratio considered in this paper, i.e., $y=0.3$,
confirmed the observation that the FFPS curves for $\Delta = 0.1$
were thermodynamically stable for intermediate size ratios. We did
not succeed in obtaining a satisfactory sampling of the Gibbs free
energy within IETs approaches for this size ratio because of
convergence problems of the numerical algorithm; thus, we just
show the results obtained within the perturbation theory applied
to the depletion potential. As shown in Figure \ref{fig. 7}, the
FFPS critical points were located above the freezing line for both
$\Delta = 0.05$ (see dotted and dot-dashed line of the left panel)
and $\Delta = 0.1$ (see right panel). We concluded that
fluid-fluid phase separation is metastable with respect to the
freezing transition for nonadditive parameters lower than $0.1$.

\section{Concluding remarks}

In this paper we calculated the fluid-fluid phase separation
equilibrium conditions of a nonadditive hard sphere mixture for a
wide range of size ratios and for two, slightly nonadditive
repulsive interactions. To estimate the coexistence lines we
evaluated the Gibbs free energy within the frameworks provided by
the modified-hypernetted-chain and the Rogers-Young integral-equation
closures of the Ornstein-Zernike equations. An alternative
approach to the phase diagram may be based on the use of
zero-separation theorems to obtain directly the chemical
potentials \cite{lee}. This method to improve the performance of
integral equations has been recently shown to be able to 
reproduce with good accuracy the phase diagram of nonadditive hard sphere mixtures in
random pores \cite{pellicane}. Furthermore, the use of zero-separation
theorems considerably lowers the computational effort when
integral-equation theories are solved in the framework of
constant-pressure calculations \cite{pastore}. On the other hand,
the application of zero-separation theorems turns out to be feasible provided
that an appropriate closure of the Ornstein-Zernike equations, such as
the modified-Verlet one, is adopted.

The modified-hypernetted-chain and the Rogers-Young coexistence curves
have been compared with the coexistence results obtained through a
first-order perturbation theory applied to the effective depletion
interaction between the larger spheres. The thermodynamic stability of
fluid-fluid phase separation with respect to freezing has been tested
as well. We found a quantitative matching between the
modified-hypernetted-chain and the Rogers-Young estimates for all the
case studies considered here. As far as the comparison between the
integral equation theories and thermodynamical perturbation
theory€ coexistence curves is concerned, the agreement is only
qualitative. However, both methods suggest that an amount of
nonadditivity as small as $\Delta \approx 0.1$ can be enough,
for intermediate size ratios, to drive the fluid-fluid critical
point into the thermodinamically stable region of the phase
diagram. The approach presented in this paper to study the phase
diagram of a nonadditive hard-sphere mixture can be easily
extended to more realistic, nonadditive models of He--rare-gas
mixtures in extreme conditions of temperature and pressure, such
as mixtures of particles interacting through modified Buckingam
potentials \cite{barrat}. Calculations in this direction are in
progress.

\clearpage

\begin{table}
\caption{Comparison between the critical molar fractions and
(total and partial) packing fractions as a function of the size
ratio and of the nonadditivity parameter for the integral
equations (IETs) and perturbation theory (PT), respectively. The
estimated error affects the second digit and its value is reported
inside quotes for the IETS only when it turns out to be greater
than $0.01$.} \label{t2}
\begin{tabular}{llccccccc} \\
%\hline & & & & &\multicolumn{2}{c} {IET} & & & & & {PT} \\
\cline{4-9}
$y$ & $\Delta$ & & & $\eta$ & $x_{1}$ & $\eta_{1}$ & $\eta_{2}$ &  \\
\hline
     &      & IETs & & 0.41 & 0.43 & 0.27 & 0.15 &     \\
     & 0.05 &      & &      &      &      &      &     \\
     &      &  PT  & & 0.42 & 0.55 & 0.31 & 0.1  &     \\
0.75 &      &      & &      &      &      &      &     \\
     &      & IETs & & 0.30 & 0.44 & 0.20 & 0.10 &     \\
     & 0.1  &      & &      &      &      &      &     \\
     &      &  PT  & & 0.30 & 0.52 & 0.22 & 0.09 &     \\
\hline
     &      & IETs & & 0.43 & 0.39 & 0.33 & 0.11 &     \\
     & 0.05 &      & &      &      &      &      &     \\
     &      &  PT  & & 0.42 & 0.43 & 0.35 & 0.08 &     \\
0.6  &      &      & &      &      &      &      &     \\
     &      & IETs & & 0.32 & 0.41 & 0.25 & 0.07 &     \\
     & 0.1  &      & &      &      &      &      &     \\
     &      &  PT  & & 0.29 & 0.32 & 0.2  & 0.1  &     \\
\hline
     &      & IETs & & 0.46 & 0.34 & 0.37 & 0.09 &     \\
     & 0.05 &      & &      &      &      &      &     \\
     &      &  PT  & & 0.43 & 0.32 & 0.35 & 0.09 &     \\
0.5  &      &      & &      &      &      &      &     \\
     &      & IETs & & 0.33 (3) & 0.34 (9) & 0.28 (4) & 0.06 (2)& \\
     & 0.1  &      & &      &      &      &      &     \\
     &      &  PT  & & 0.32 & 0.26 & 0.23 & 0.08 &     \\
\hline
\end{tabular}
\end{table}

\clearpage
\newpage

\begin{figure}
\begin{center}
\includegraphics[width=9cm,angle=-90]{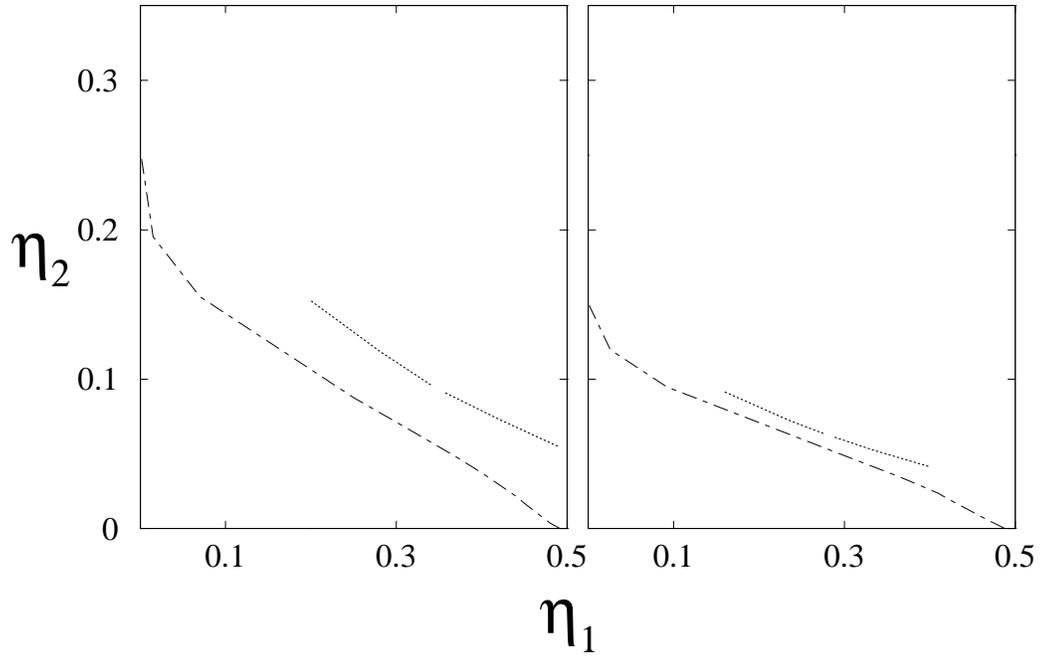}
\end{center}
\caption{Fluid-fluid and fluid-solid transition lines in the
$\eta_2$ vs. $\eta_1$ representation for $y=0.3$, $\Delta=0.05$
(left panel) and $\Delta=0.1$ (right panel). See Figure
 \ref{fig. 4} for the legend. At variance with the previous
figures we do not report the IET coexistence lines.} \label{fig.
7}
\end{figure}

\end{document}